\documentclass{ws-ijmpe}
\usepackage[super,compress]{cite}
\usepackage{epstopdf}
\usepackage{epsfig}
\usepackage{comment}
\begin{document}

\markboth{M. EL ADRI $\&$ M. OULNE}{}

\catchline{}{}{}{}{}

\title{Shell Evolution in Neutron-rich Ge, Se, Kr and Sr Nuclei within RHB Approach}

\author{M. EL ADRI\footnote {corresponding author}  and M. OULNE$^\dagger$.}

\address{High Energy Physics and Astrophysics Laboratory, Department of Physics, \\Faculty of Sciences SEMLALIA, Cadi Ayyad University,  \\P.O.B. 2390, Marrakesh, Morocco.\\
$^*$mohamed.eladri@edu.uca.ma\\
$^\dagger$oulne@uca.ma}

\maketitle

\begin{history}
\received{Day Month Year}
\revised{Day Month Year}
\end{history}

\begin{abstract}
	The exotic even-even isotopic chains from Z=32 to Z=38 are investigated by means of the relativistic Hartree-Bogoliubov (RHB) approach with the explicit Density Dependent Meson-Exchange (DD-ME2) and Density Dependent Point-Coupling (DD-PC1) models. The classic magic number N=50 is reproduced and the new number N=70 is predicted to be a robust shell closure by analysing several calculated quantities such as:  two-neutron separation energies, two-neutron shell gap, neutron pairing energy, potential energy surface and neutron single particle energies. The obtained results are compared with the predictions of finite range droplet model (FRDM) and with the available experimental data. A reasonable and satisfactory agreement between the theoretical models and experiment is established.
\end{abstract}


\ccode{PACS numbers: 21.10-k, 21.10.Dr, 21.10.Ft, 21.60-n}


\section{introduction}
    The recent generation of radioactive beam facilities allows us to study nuclei so far from the valley of stability. Therefore, a reliable study of such nuclei requires a consistent theory that can describe nuclei in both ground and excited states. In this context, many nuclear approaches have emerged. Among these, the ab-initio  \cite{Puldiner,Hagen,Barrett,Drut} and shell-model methods which are successfully used in description of light nuclei. However, the complexity of calculations has limited the application of ab-initio method in medium, heavy and super-heavy ones, whereas significant progress has been reported in the development of shell-model techniques which use sophisticated truncation schemes \cite{Horoi} to predict the  properties of heavier nuclei. Nowadays, with the mean-field concept, the microscopic approaches based on nuclear Energy Density Functionals (EDFs) are the most accurate tools that can be applied in all over the nuclear chart including those very close to  drip lines. Generally, there are three classes of nuclear EDFs: the Skyrme energy functional \cite{Vautherin1969,Vautherin1972}, the Gogny effective interaction \cite{Gogny} and the relativistic mean-field model \cite{Walecka1974,Boguta1977}, which are widely used today and capable to compete with the micro-macroscopic method on a quantitative level. 
    
    Recently, many studies indicate that Covariant Density Functional Theory (CDFT) is quite reliable and accurate in analysing and describing the nuclear structure, not only in nuclei lying at the valley of stability, but also in those located far from it, which are usually dubbed "exotic nuclei", as it has been proved in Refs \cite{Niksic2011,Agbemava2015,Bassem}. In 2016, J. Lie and al. \cite{Lie} have studied the formation of new shell gaps in intermediate mass neutron-rich nuclei within the relativistic Hartree-Fock-Bogoliubov (RHFB) theory, in which they successfully reproduced the well-known shell gaps and suggested the occurrence of new ones. In our recent work \cite{ELADRI}, we have confirmed the emergence of new magic numbers at N=32 and N=40 in Ca (Z=20) and Ar (Z=18) nuclei within the relativistic Hartree-Bogoliubov (RHB) and non-relativistic Hartree-Fock-Bogoliubov (HFB) approaches. The classic magic numbers in those nuclei are also well reproduced. Furthermore, the neutron shell gap at N=70 for neutron-rich Ni (Z=28) nuclei has been predicted in M. Bhattacharya's review \cite{Bhattacharya} within the relativistic mean field (RMF). The persistence of the closed neutron shell at N=50 from Kr (Z=36) all the way down to Ga (Z=31) as well as in Ni (Z=28) is also suggested in Ref. \cite{Hakala}.
       
     It is worth to note that both experimental and theoretical studies have recently shown that the shell gap can be changed locally in exotic nuclei. Indeed, the well-known magic numbers for nuclei located in the valley of stability or very close to it can disappear and new ones can appear instead. Examples for experimental studies can be found in Refs \cite{Nakamura,Motobayashi}. For theoretical investigations, in addition to the references quoted in the previous paragraph, see for instance \cite{Kanungo,Grawe}. In this context, we aim to study and analyse the evolution of the shell gap of neutron rich-nuclei. To this end, we have selected four isotope chains from Z=32 to Z=38 covering the region of neutron numbers $42\leq N\leq76$. Of course, this choice was not arbitrary, but rather based on the study of the entire nuclear chart by using the relativistic Hartree-Bogoliubov (RHB) approach with DD-PC1 \cite{Niksic2008,Lalazissis2005} and DD-ME2 \cite{Niksic2014} parametrizations. As a result, we found that only the four selected nuclei show magicity feature at N=70 in the exotic region.
 	
    In the present work, we employ the relativistic EDFs, also called Covariant Density Functional Theory (CDFT), with two classes, namely: Density Dependent Meson-Exchange (DD-ME) and Density Dependent Point-Coupling (DD-PC) parameters in order to investigate the shell evolution in the exotic region. The article is organized as follows: Sect \ref{Sec2} shortly outlines the approach that we have used to do our calculations. In sect \ref{Sec3} numerical tests as well as the input details and the interactions used in calculations are presented. The obtained results are analysed and discussed in Sect \ref{Sec4}. Finally, the main conclusions are given in Sect \ref{Sec5}.
 	
 \section{Theoretical framework} \label{Sec2}
 Covariant Density Functional Theory (CDFT), also often labelled as relativistic Hartree-Bogoliubov (RHB) theory, is a microscopic theoretical tool that can be used to describe the entire nuclear chart with success. In the present work, we have employed two classes of relativistic Hartree-Bogoliubov theories. The first one is the DD-PC model \cite{Niksic2008,Lalazissis2005} which is characterised by a zero-range interaction and the second one is the DD-ME model \cite{Niksic2014} which uses a finite interaction range. A brief description of these models is given in the following subsections.
 \subsection{The Density-Dependent Meson-Exchange\\}\label{Sub1}
 The basic building blocks of relativistic Hartree-Bogoliubov for DD-ME is the standard Lagrangian density with medium dependent vertice \cite{Niksic2014}
 \begin{eqnarray}
 \mathcal{L}  &  =\bar{\psi}\left[
 \gamma(i\partial-g_{\omega}\omega-g_{\rho
 }\vec{\rho}\,\vec{\tau}-eA)-m-g_{\sigma}\sigma\right]  \psi\nonumber\\
 &  +\frac{1}{2}(\partial\sigma)^{2}-\frac{1}{2}m_{\sigma}^{2}\sigma^{2}%
 -\frac{1}{4}\Omega_{\mu\nu}\Omega^{\mu\nu}+\frac{1}{2}m_{\omega}^{2}\omega
 ^{2}\\\nonumber
 &  -\frac{1}{4}{\vec{R}}_{\mu\nu}{\vec{R}}^{\mu\nu}+\frac{1}{2}m_{\rho}%
 ^{2}\vec{\rho}^{\,2}-\frac{1}{4}F_{\mu\nu}F^{\mu\nu}\nonumber
 \label{lagrangian}
 \end{eqnarray}
 with $\psi$ is Dirac spinor and $m$ is the bare nucleon mass. $m_{\sigma}$, $m_{\omega}$, $m_{\delta}$ and $m_{\rho}$ are meson masses.  $g_{\sigma}$, $g_{\omega}$, $g_{\delta}$ and $g_{\rho}$ are the coupling constants and e corresponds to the proton's charge. It vanishes for neutron. $\Omega_{\mu\nu}$, ${\vec{R}}^{\mu\nu}$, $F_{\mu\nu}$ denote fields tensors.\\
 \begin{eqnarray}
 \Omega_{\mu\nu}=\partial^{\mu}\Omega^{\nu}-\partial^{\nu}\Omega^{\mu}
 \end{eqnarray}
 \begin{eqnarray}
 \vec{R}^{\mu\nu}=\partial^{\mu}\vec{\rho}^{  \nu}-\partial^{\nu}\vec{\rho}^{ \mu}
 \end{eqnarray}
 \begin{eqnarray}
 F_{\mu\nu}=\partial^{\mu}A^{\nu}-\partial^{\nu}A^{\mu}
 \end{eqnarray}
 This linear model that was established by Walecka \cite{Walker1986,Walecka1974} failed to provide a quantitative description of nuclear system. For instance, it gives a quit large compressibility coefficient of nuclear matter, about 500 MeV \cite{Boguta1977}. Therefore, to describe a complex nuclear system properties,  a non-linear potential has been added to the Lagrangian, and replaces the $\frac{1}{2}m^2_{\sigma}\sigma^2$ term, namely
 \begin{equation}
 U_(\sigma)=\frac{1}{2}m^2_{\sigma}\sigma^2+\frac{1}{2}g_{2}\sigma^3+\frac{1}{2}g_{3}\sigma^4
 \end{equation}
 Using both DD-ME and DD-PC parametrizations, many studies proved that this model can reproduce the experimental data with high precision \cite{Lalazissis2005,Niksic2008,Typel1999}.
 
 The coupling of the $\sigma$  and $\omega$ mesons  to the nucleon field reads \cite{Niksic2014}
 \begin{equation}
 g_i(\rho)=g_i(\rho_{sat})f_i(x) \qquad for  \quad i=\sigma,  \omega
 \end{equation}
 with the density dependence  given by
 \begin{equation}
 \label{dep}
 f_i(x)=a_i\frac{1+b_i(x+d_i)^2}{1+c_i(x+d_i)^2}
 \end{equation}
 where $x=\rho/\rho_{sat}$, $\rho$ is the baryonic density and $\rho_{sat}$ is the baryon density at saturation in symmetric nuclear matter.
 In eq (\ref{dep}), the parameters are not independent, but constrained as follows:
 $f_i(1)=1$, $f''_{\sigma}(1)=f''_{\omega}(1)$, and $f''_i(0)=0$.
 These constraints reduce the number of independent parameters for the density dependence.\\
 In the $\rho$-meson case, we have an exponential density dependence
 \begin{equation}
 g_{\rho}(\rho)=g_{\rho}(\rho_{sat})e^{-a_\rho(x-1)} 
 \end{equation}
 The isovector channel is parametrized by $g_{\rho}(\rho_{sat})$ and $a_\rho$

 \subsection{The Density-Dependent Point-Coupling\\}
 The effective Lagrangian density of DD-PC model is defined by \cite{Niksic2008}
 \begin{eqnarray}\label{LagDDPC}
 \mathcal{L}& =\bar{\psi}(i\gamma.\partial-m)\psi
  -\frac{1}{2}\alpha_{S}(\rho)(\bar{\psi}\psi)(\bar{\psi}\psi)\\\nonumber
  &
  -\frac{1}{2}\alpha_{V}(\rho)(\bar{\psi}\gamma^{\mu}\psi)(\bar{\psi}\gamma_{\mu}\psi) -\frac{1}{2}\alpha_{TV}(\rho)(\bar{\psi}\vec{\tau}\gamma^{\mu}\psi)(\bar{\psi}\vec{\tau}\gamma_{\mu}\psi)\\\nonumber
  &
  -\frac{1}{2}\delta_S(\partial_{\nu}\bar{\psi}\psi)(\partial_{\nu}\bar{\psi}\psi)-e\bar{\psi}\gamma.A.\frac{1-\tau_3}{2}\psi\\\nonumber
 \end{eqnarray}
 This Lagrangian contains the isoscalar-scalar interaction ($\sigma$ meson) $(\bar{\psi}\psi)(\bar{\psi}\psi)$, isoscalar-vector interaction ($\omega$ meson) $(\bar{\psi}\gamma^{\mu}\psi)(\bar{\psi}\gamma_{\mu}\psi)$, isovector-vector interaction ($\rho $ meson) $(\bar{\psi}\vec{\tau}\gamma^{\mu}\psi)(\bar{\psi}\vec{\tau}\gamma_{\mu}\psi)$ and their corresponding gradient couplings $\partial_{\nu}(...)\partial^{\nu}(...)$. 
 It also contains the free-nucleon Lagrangian, the point-coupling interaction terms and the coupling of protons to the electromagnetic field.
 The derivative terms in eq. (\ref{LagDDPC}) account for the main effects of finite range interactions which are important for a quantitative description of nuclear density distribution.\\
 The functional form of the couplings is given by
 \begin{equation}
 \alpha_i(\rho)=a_i+(b_i+c_ix)e^{-d_ix} \qquad for \quad i=S,T,TV
 \end{equation}
 where $x=\rho/\rho_{sat}$ and $\rho_{sat}$ denotes the nucleon density in units of the saturation density of symmetric nuclear matter.
 For more details see Ref \cite{Niksic2014} 
 \section{Numerical Details}\label{Sec3}
 This work is realized by using the relativistic Hartree-Bogoliubov (RHB) theory based on the DD-ME2 and DD-PC1 parametrizations and separable pairing within the DIRHB computer code \cite{Niksic2014}, in which the RHB equations can be solved iteratively in a basis of spherical, axially symmetric or triaxial harmonic oscillator (HO). \\
 
 In the present work, we have used the finite range pairing interaction separable in coordinate space which was proposed by Tian et al. \cite{Tian2009}. It is given in the pp-channel by 
 \begin{equation}
 V^{pp}(\mathbf{r_1,r_2,r'_1,r'_2})=-G\delta(\mathbf{R-R'})P(r)P(r'),
 \end{equation}
 Where $\mathbf{R}=\frac{1}{2}(\mathbf{r_1+r_2})$ is the centre of mass, $\mathbf{r=r_1-r_2}$ are the relative coordinates and $P(r)$ represents the form factor which is given by
 \begin{equation}
 P(r)=\frac{1}{(4\pi a^2)^{3/2}}e^{-\frac{r^2}{4a^2}}
 \end{equation} 
 The two parameters: the pairing strength $G$ and the pairing width $a$  have been adjusted to reproduce the density dependence of the gap at the Fermi surface. The following values: $G=728 MeV.fm^3$ and $a=0.6442 fm$ which were determined for the D1S parametrization \cite{Tian2009} of the Gogny force have also been used here.
  The numbers of Gauss-Laguerre $N_{GL}$ and Gauss-Hermite $N_{GH}$ mesh-points were $N_{GL}$=$N_{GH}$=48, and the number of Gauss-Legendre mesh-points was $N_{GLEG}=80$.
 
\begin{figure}[tbh]
 			\centering \includegraphics[scale=0.45]{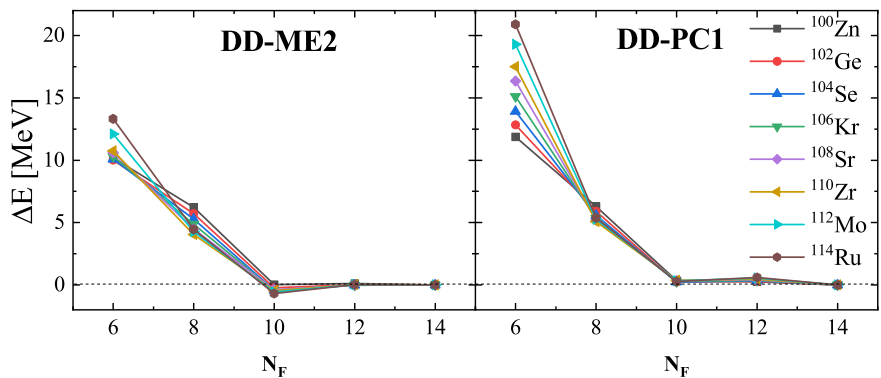}
 	\caption{Binding Energy difference between a reference
 		calculation performed with 14 shells and calculations performed with $N_{F}$ shells for the N=70 isotones ($30 \le Z \le 44$) by using the DD-ME2 and DD-PC1 parametrizations.\label{DivDir}}
 \end{figure}

 In order to study the convergence of the RHB results in nuclei under investigation, we have calculated the total binding energy as functions of the shells number for the fermions $N_{F}$ for the neutron-rich $^{100}Zn$, $^{102}Ge$, $^{104}Se$, $^{106}Kr$, $^{108}Sr$, $^{110}Zr$, $^{112}Mo$ and $^{114}Ru$ nuclei (N=70 isotones). As one can see in Fig. \ref{DivDir}, the total binding energy converges exactly at $N_{F}=10$. That is why all calculations performed with the DIRHB code are carried out in a safe full spherical basis of $N_{F}=10$. For the bosons, the number of shells is fixed to $N_B=20$. The $\beta_2$-deformation parameter for the harmonic oscillator basis as well as for the initial Woods-Saxon potential is set to 0, except for constrained calculations of the potential energy surface where the $\beta_2$ parameter varies in a range from -0.5 to 0.5 in steps of 0.05. The parameters of DD-ME2 and DD-PC1 relativistic energy functionals are given in Table \ref{table 1}
 \begin{table}[h]
 	\caption{The parameters of DD-ME2 \cite{Lalazissis2005} and DD-PC1 \cite{Niksic2008} interactions. The masses are given in MeV and all other parameters are dimensionless.\label{table 1}}
 	\begin{center}
 		{\begin{tabular}{@{}ccc@{\hspace{18pt}}@{\hspace{18pt}}c@{}}    \hline
 				Parameter     & DD-ME2 \cite{Lalazissis2005}      &  Parameter   & DD-PC1 \cite{Niksic2008}  \\ \hline
 				m             &       939     & m            & 939      \\
 				$m_{\sigma}$  &       550.124 & $a_{\sigma}$ & -10.04616\\
 				$m_{\omega}$  &       783.000 & $b_{\sigma}$ & -9.15042 \\
 				$m_{\rho}$    &       763.00  & $c_{\sigma}$ & -6.42729 \\
 				$m_{\delta}$  &       0.000   & $d_{\sigma}$ & 1.37235  \\
 				$g_{\sigma}$  &       10.5396 & $a_{\omega}$ & 5.91946  \\
 				$g_{\omega}$  &       13.0189 & $b_{\omega}$ & 8.86370  \\
 				$g_{\rho}$    &       3.6836  & $b_{\rho}$   & 1.83595  \\
 				$g_{\delta}$  &       0.000   & $d_{\rho}$   & 0.64025  \\
 				$a_{\sigma}$  &       1.3881  &              &          \\
 				$b_{\sigma}$  &       1.0943  &              &          \\
 				$c_{\sigma}$  &       1.7057  &              &          \\
 				$d_{\sigma}$  &       0.4421  &              &          \\
 				$e_{\sigma}$  &       0.4421  &              &          \\
 				$a_{\omega}$  &       1.3892  &              &          \\
 				$b_{\omega}$  &       0.9240  &              &          \\
 				$c_{\omega}$  &       1.4620  &              &          \\
 				$d_{\omega}$  &       0.4775  &              &          \\
 				$e_{\omega}$  &       0.4775  &              &          \\
 				$a_{\rho}$    &       0.5647  &              &          \\ \hline
 		\end{tabular}}
 	\end{center}
 \end{table}
 \section{Results and Discussion}\label{Sec4}
 In this section, we present and analyse results for the even-even isotopes $_{32}Ge$, $_{34}Se$, $_{36}Kr$ and $_{38}Sr$. By calculating several observables such as Fermi levels, two-neutron separation energies $S_{2n}$, two-neutron shell gap $\delta_{2n}$, neutron pairing energy $E_{pair}$, potential energy surface (PES) and neutron single particle energies (SPE), we discuss the evolution of the neutron shell closure at N=50 and N=70. In all cases, the obtained results  within the RHB approach based on DD-ME2 \cite{Niksic2014} and DD-PC1 \cite{Niksic2008,Lalazissis2005} parametrizations are compared with the available experimental data \cite{Wang} and with the predictions of micro-macroscopic model FRDM \cite{Moller}.
 
 \subsection{Fermi Levels}
\begin{figure}[tbp]
 		\centering \includegraphics[scale=0.45]{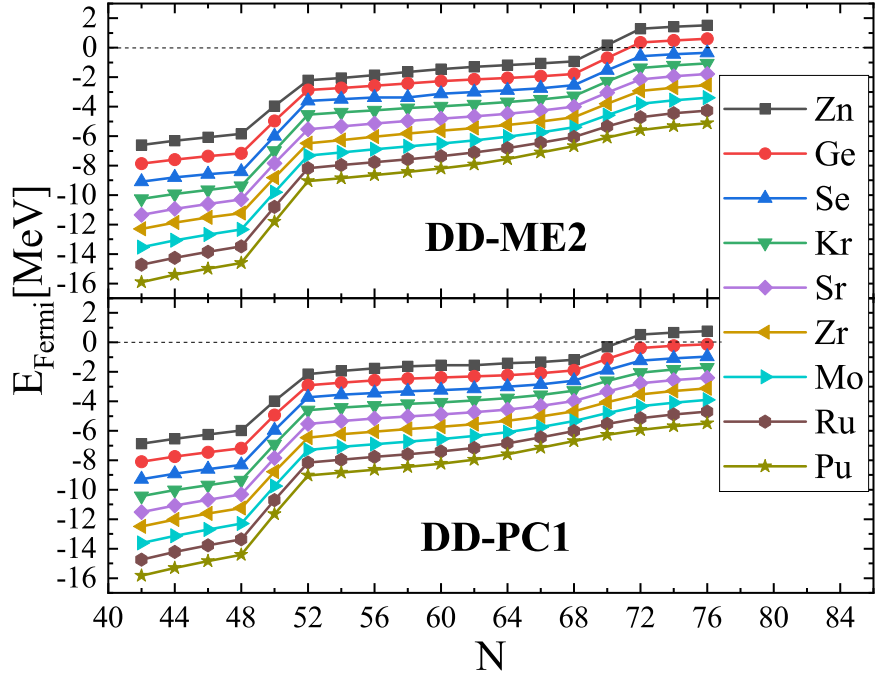}  
 	\caption{(Color online) The calculated Fermi levels in even-even isotopes with $(30 \le Z \le 44)$ using the DD-PC1 and the DD-ME2 sets.}\label{EfFig}
 \end{figure}

 As pointed out in the introduction, the choice of studied nuclei, namely: Ge, Se, Kr and Sr, was not random but was based on the study of all nuclei throughout the periodic table. The analysis has shown that N=70 did not behave as a strong magic number expected in the selected ones. Fig. \ref{EfFig} exhibits the Fermi levels of the selected nuclei as well as those laying in their vicinity around N=70 (i.e. N=70 isotones). In all these nuclei, the well-known magic number N=50 has been reproduced by both models (DD-ME2 and DD-PC1), as there is an abrupt jump in the Fermi levels. The same trend is observed around N=70 except for some nuclei. In Ge, Se, Kr and Sr nuclei, the jumps in Fermi levels  strongly support N=70 as a shell closure for both  functionals (DD-ME2 and DD-PC1). Above Sr nucleus,  this gap becomes more and more weak with increasing protons number Z until quenching in agreement with the results of Ref \cite{Bender}. This is more clear with the DD-PC1 expectations. Below Ge nucleus, the relative position of the neutron drip line is located under  N=70, as shown in Fig. \ref{EfFig}. Due to all these considerations, we have restricted our study of the neutron shell closure evolution at N=70 to the four indicated nuclei (Ge, Se, Kr and Sr) which are investigated in the following sections using a variety of quantities to confirm and support this new candidate shell closure.
 
 \subsection{Two-neutron separation energies }\label{S2nSec}
\begin{figure}[tbh]
 \begin{center}
 	\begin{minipage}[]{0.5\textwidth}
 		\centering \includegraphics[scale=0.45]{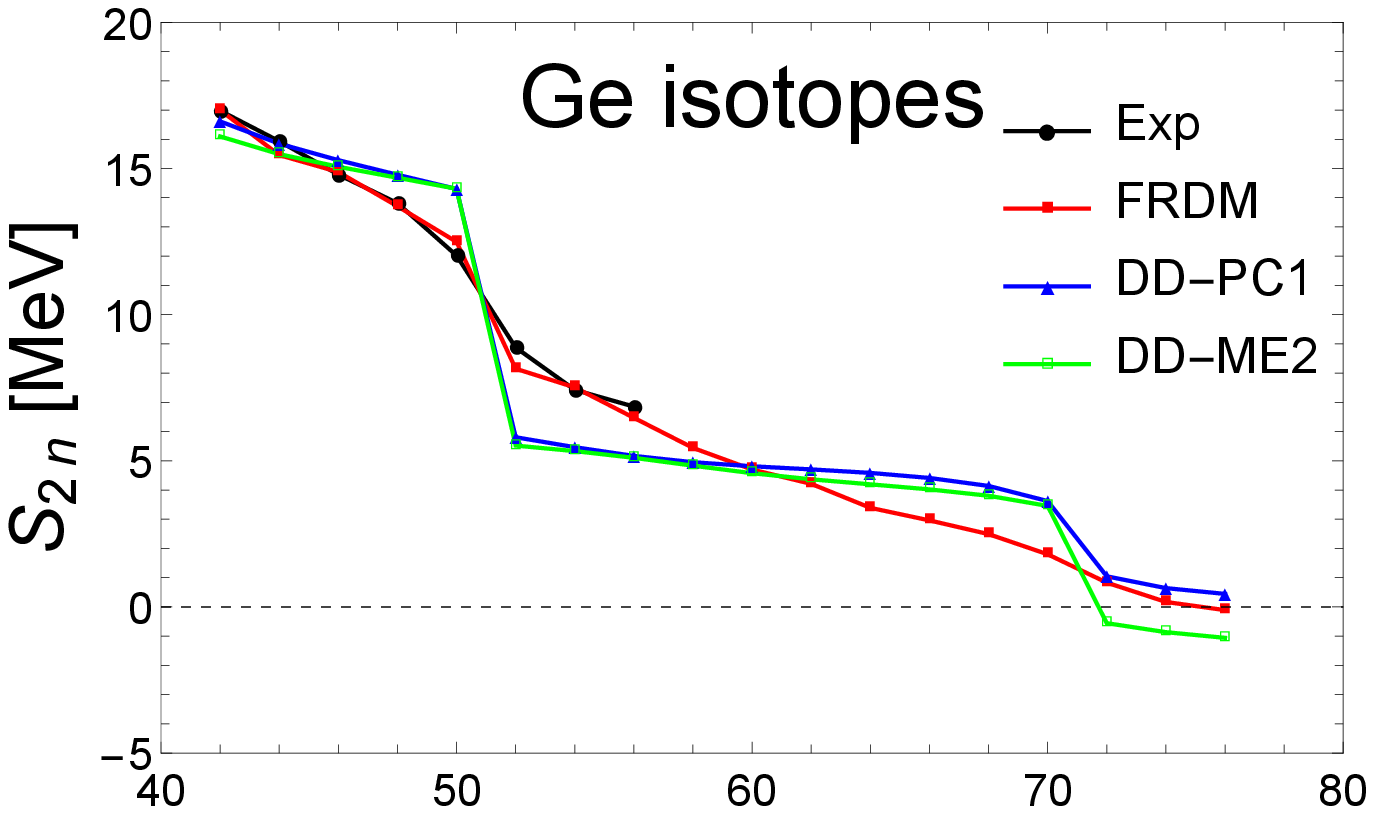}
 	\end{minipage}\hfill
 	\begin{minipage}[]{0.5\textwidth}
 		\centering \includegraphics[scale=0.45]{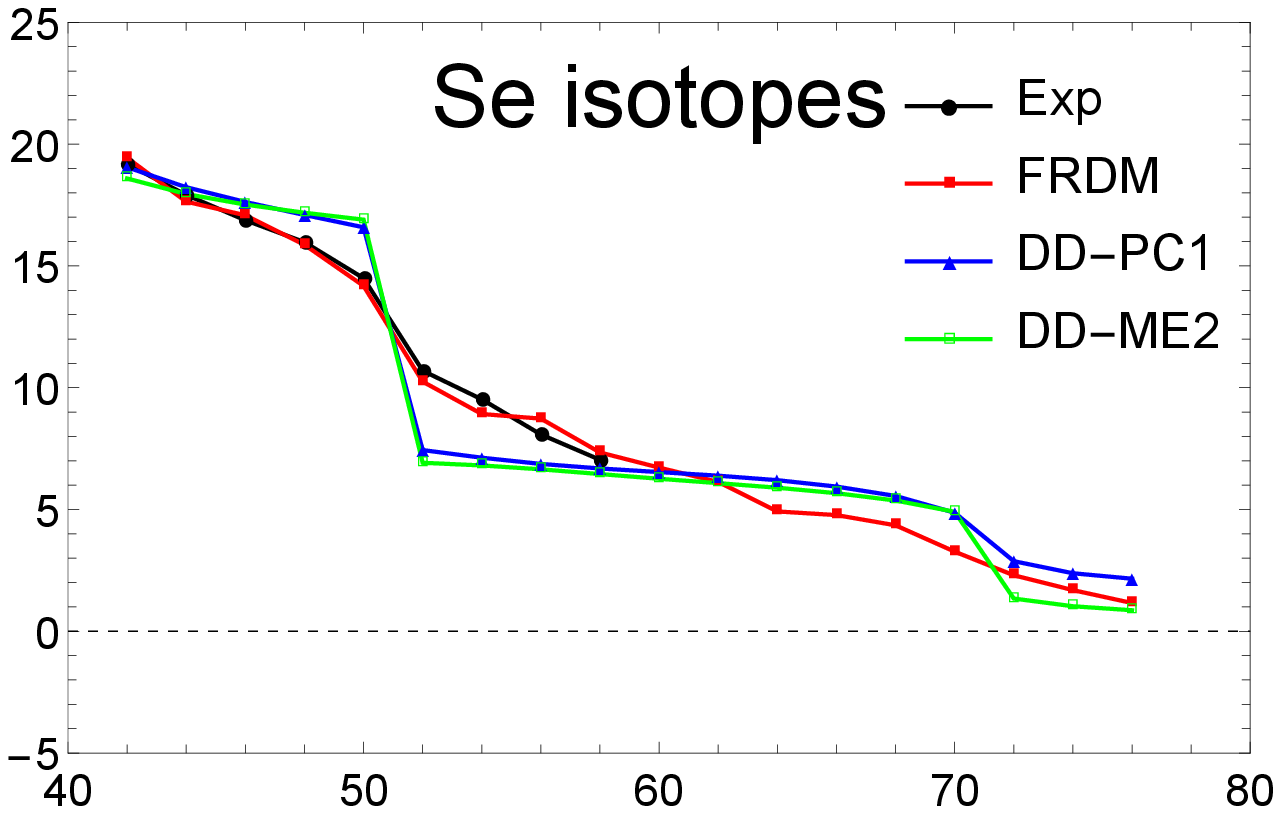}
 	\end{minipage}\hfill
 	\begin{minipage}[]{0.5\textwidth}
 		\centering \includegraphics[scale=0.45]{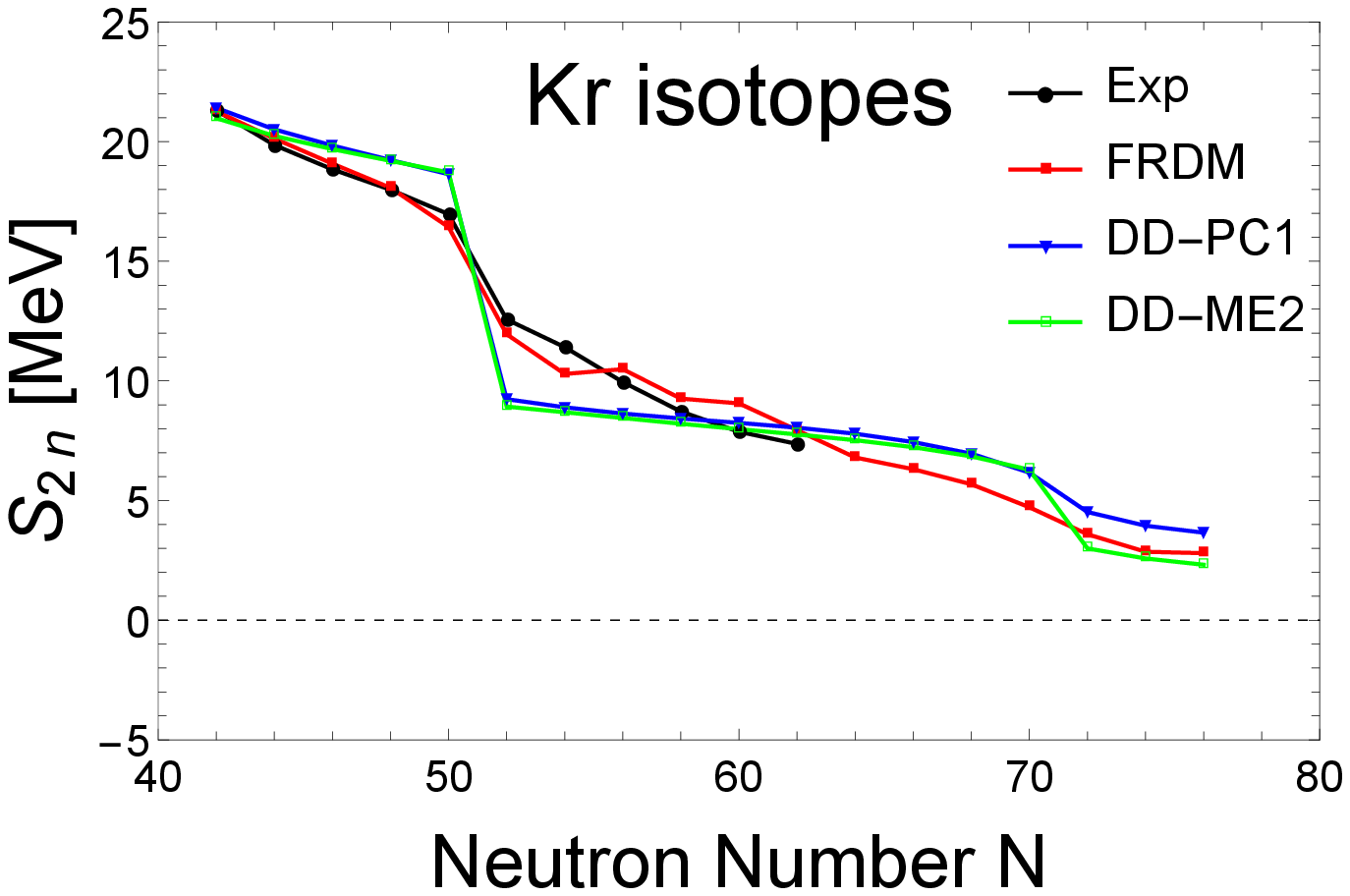}
 	\end{minipage}\hfill
 	\begin{minipage}[]{0.5\textwidth}
 		\centering \includegraphics[scale=0.45]{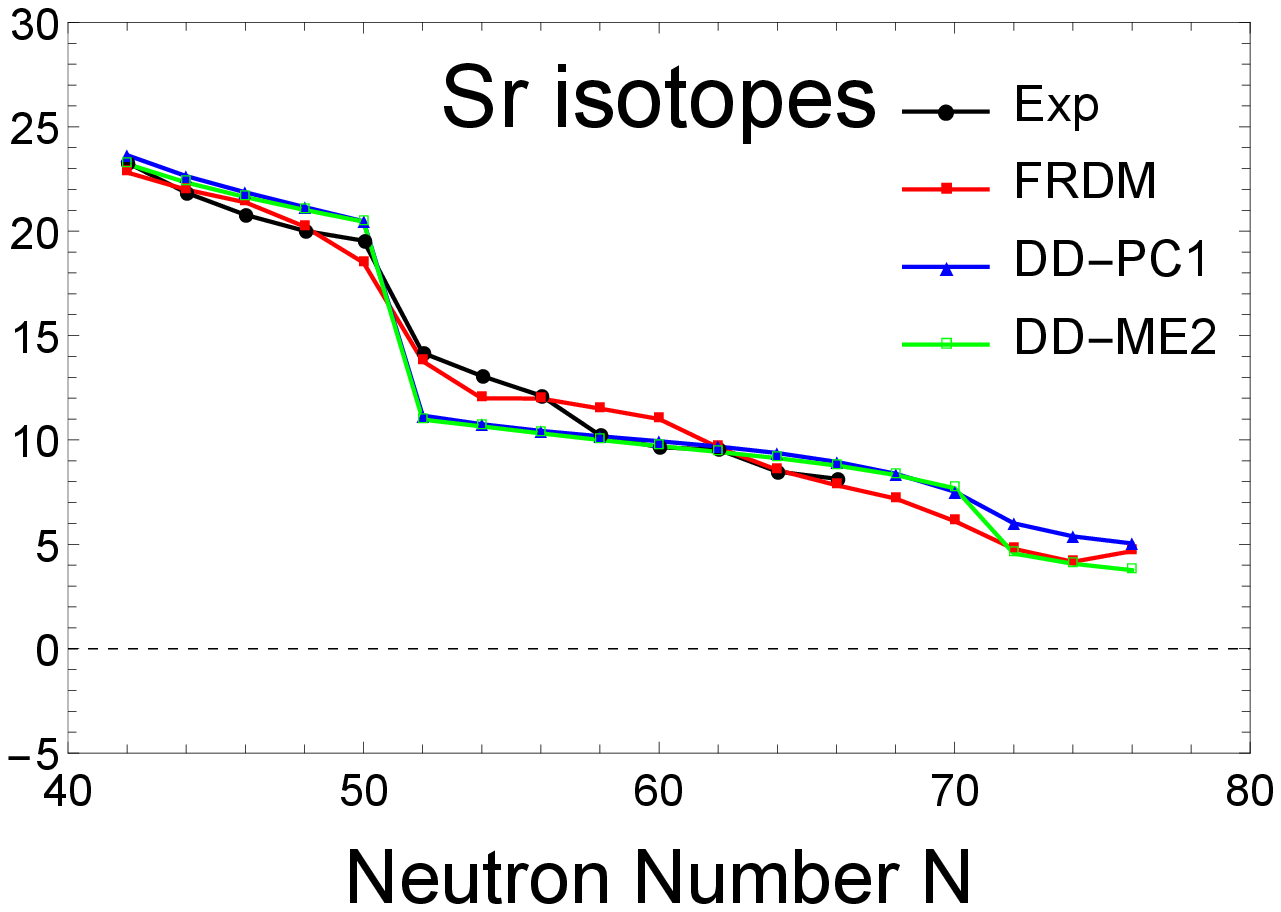}
 	\end{minipage}\hfill
 	\caption{Two-neutron separation energies as a function of neutron number N for  Ge, Se, Kr and Sr, obtained by RHB with DD-PC1 and DD-ME2 parametrizations, and compared with FRDM predictions and available experimental data.\label{S2nFig}}
 \end{center}
 \end{figure}

 The two-neutron separation energy is considered as an essential physical quantity in exploring the shell closures. It is given by 
 \begin{equation}
 S_{2n}(Z,N)=BE(Z,N)-BE(Z,N-2)
 \end{equation}
 where $BE(Z,N)$ is the positive value of binding energy of a given nucleus with $Z$ protons and $N$ neutrons.\\
 For the even-even $_{32}Ge$, $_{34}Se$, $_{36}Kr$ and $_{38}Sr$ nuclei, the variations of two-neutron separation energy $S_{2n}$ with the neutron number $N$, obtained in RHB calculations where DD-ME2 and DD-PC1 parametrizations have been used, are displayed in Fig. \ref{S2nFig}, in comparison with the
 available experimental data taken from Ref \cite{Wang} and with the predictions of FRDM \cite{Moller}. From this figure, one can see that $S_{2n}$ decreases gradually with increasing N, and a remarkable drop is clearly seen in the well-known classic magic number N=50 for all investigated nuclei in both theoretical and experimental curves. In the exotic region, the same behaviour is observed at N=70 with the DD-ME2 and DD-PC1 calculations, suggesting it as a magic number. However, FRDM predictions do not show any such magic number at this location. Also, it is clear that the kink observed with DD-ME2 is more significant than that observed with DD-PC1, especially for Kr and Sr isotopes.
   
  Note that local drops in separation energies are not only due to the appearance of new magic numbers, but sometimes to other causes, such as competition and mixing of low-lying prolate and oblate shapes \cite{Vretenar}. Thus, a study using other quantities that give more information must be utilized. Pairing effect has often been used to shed more light on the shell gap.
 
 \subsection{Two-neutron shell gap}
\begin{figure}[tbh]
 	\begin{minipage}[b]{0.5\textwidth}
 		\centering \includegraphics[scale=0.45]{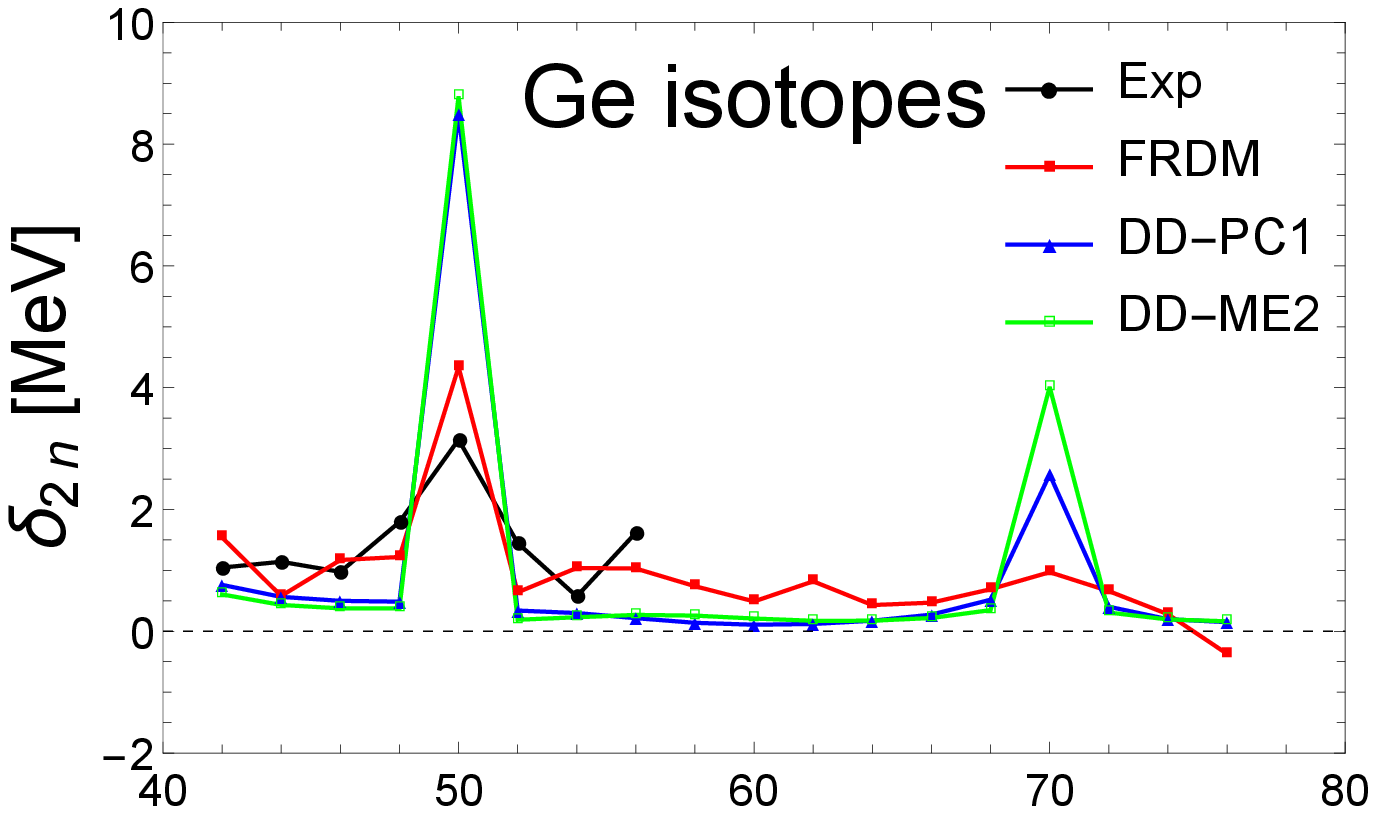}
 	\end{minipage}\hfill
 	\begin{minipage}[b]{0.5\textwidth}
 		\centering \includegraphics[scale=0.45]{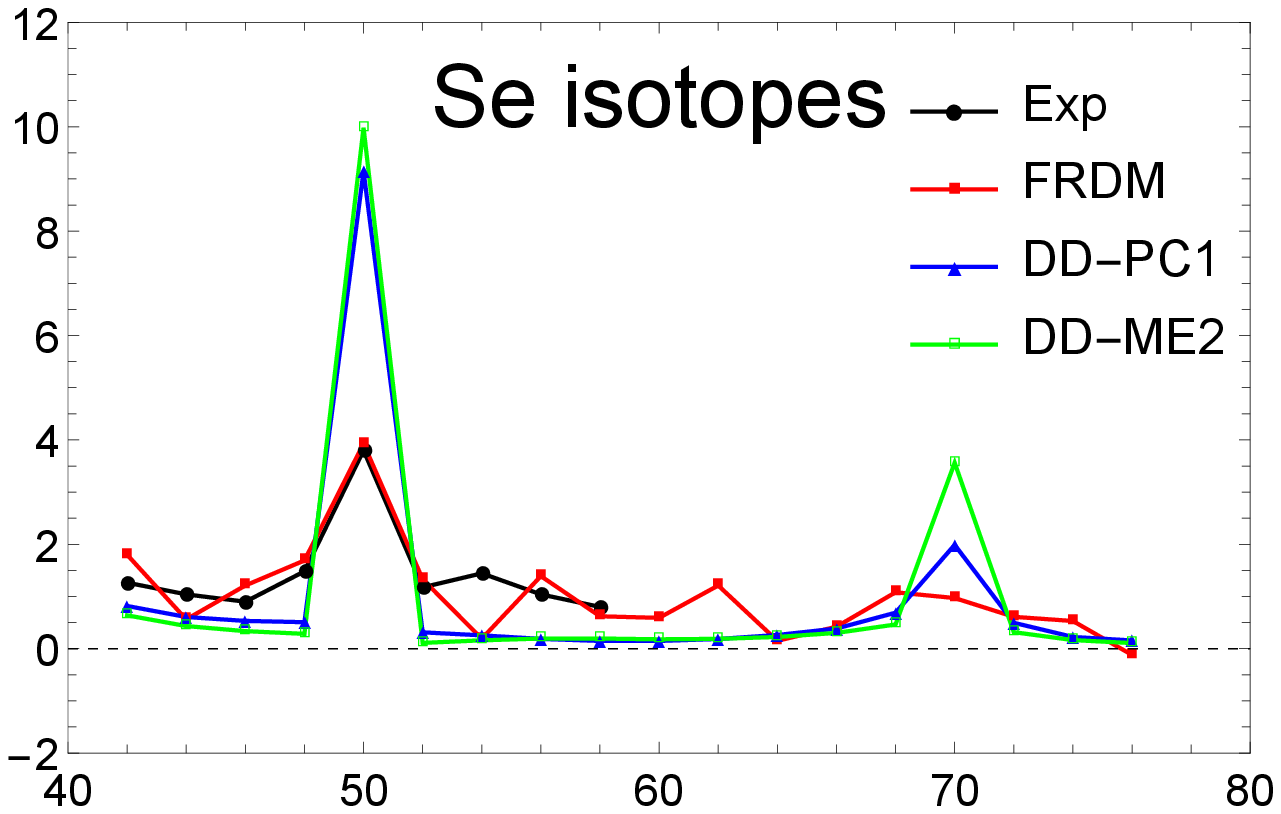}
 	\end{minipage}\hfill	
 	\begin{minipage}[b]{0.5\textwidth}
 		\centering \includegraphics[scale=0.45]{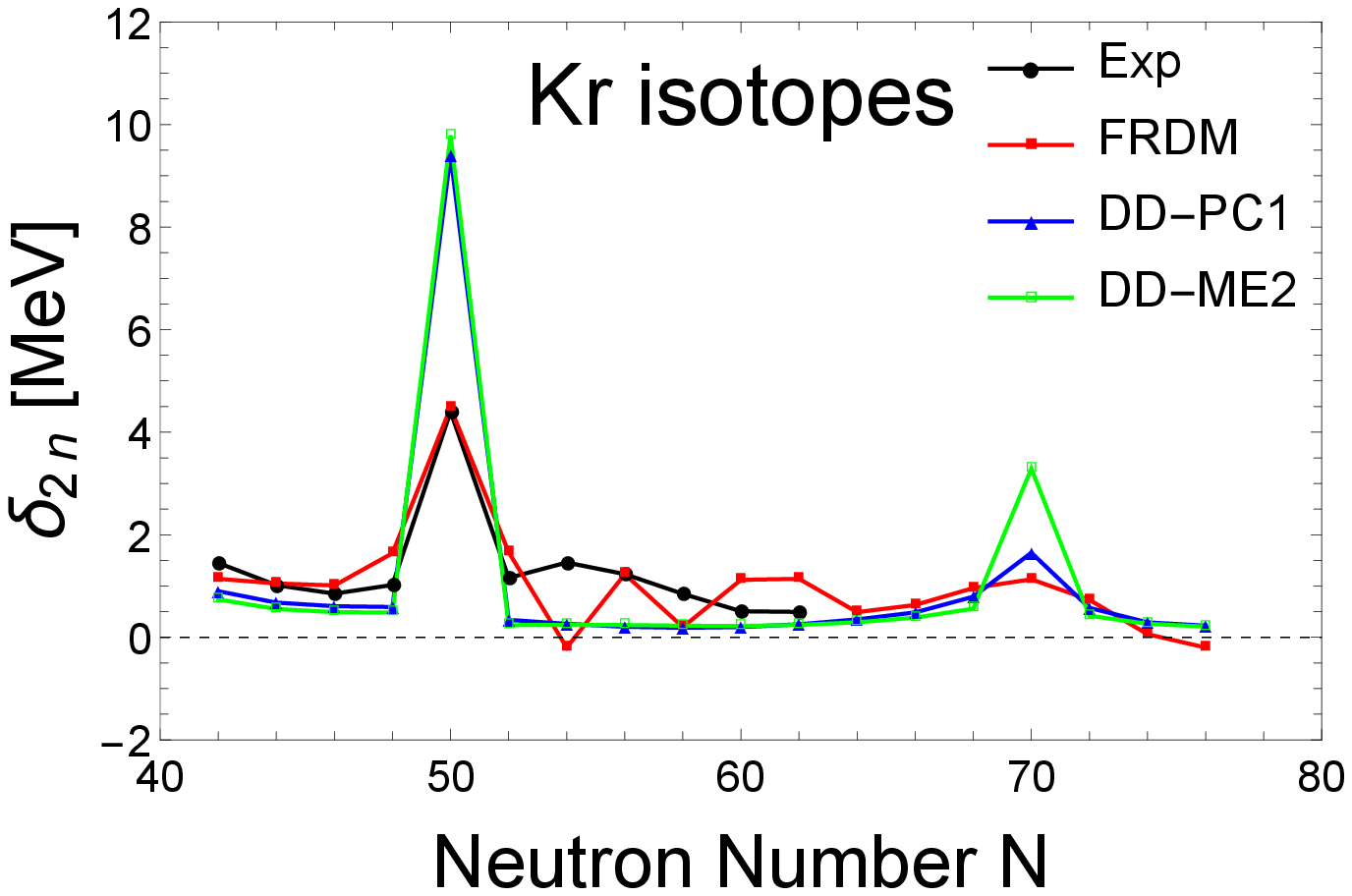}
 	\end{minipage}\hfill
 	\begin{minipage}[b]{0.5\textwidth}
 		\centering \includegraphics[scale=0.45]{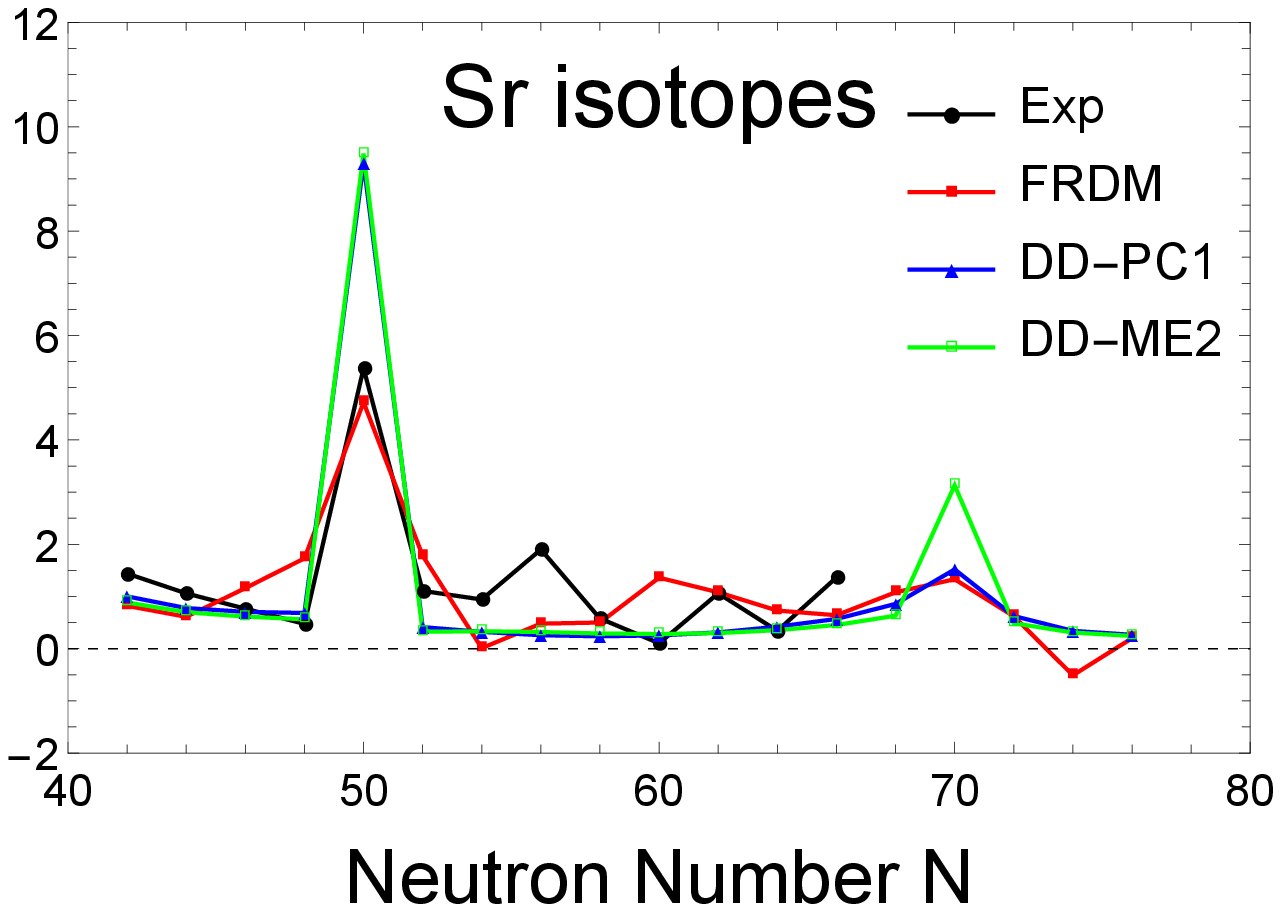}
 	\end{minipage}\hfill	
 	\caption{The two-neutron shell gap $\delta_{2n}$, for Ge, Se, Kr and Sr obtained by RHB with DD-PC1 and DD-ME2 sets, and compared with FRDM predictions and the available experimental data.}\label{D2nFig}
 \end{figure}

  We now discuss two-neutron shell gap $\delta_{2n}$ that is given by
  	\begin{eqnarray}
  				\delta_{2n}=S_{2n}(Z,N)-S_{2n}(Z,N+2)
  	\end{eqnarray}
  This entity is very sensitive and efficient in detecting shell closures. It presents an intense pick when it crosses a magic number. Fig. \ref{D2nFig} exhibits $\delta_{2n}$ as a function of neutron number N for the four investigated nuclei obtained by the RHB approach in comparison with FRDM model and with the available experimental data. The overall agreement with experiment is excellent. 
  
  As can be seen in Fig. \ref{D2nFig}, the theoretical and experimental results show a sharp peak in $\delta_{2n}$ at the classic magic number N=50. However, in the exotic region, there is a discrepancy between the theoretical approaches. The microscopic models (DD-ME2 and DD-PC1) suggest  N=70 to be a neutron shell closure, while, the micro-macroscopic model (FRDM) shows a weak gap at this shell. Besides, DD-ME2 presents a sharp peak with respect to DD-PC1. This remark is more clear in Kr and Sr isotopes. In addition, FRDM model shows a small kink at certain neutron numbers (N=58, 62 and 64) while our calculations show no change at these positions.
  
 \subsection{Neutron pairing energy}
\begin{figure}[tbh]
 	\begin{minipage}{0.5\textwidth}
 		\centering \includegraphics[scale=0.45]{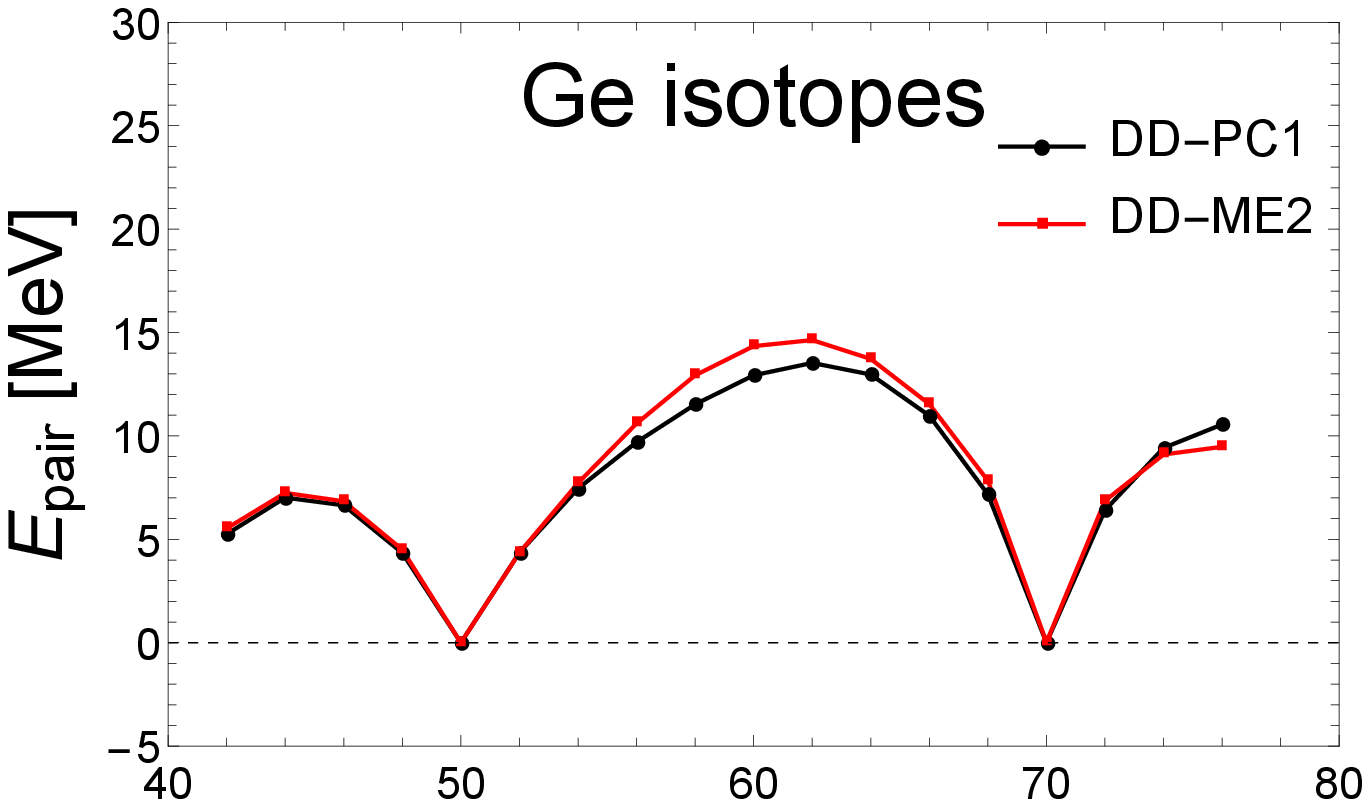}
 	\end{minipage}\hfill
 	\begin{minipage}{0.5\textwidth}
 		\centering \includegraphics[scale=0.45]{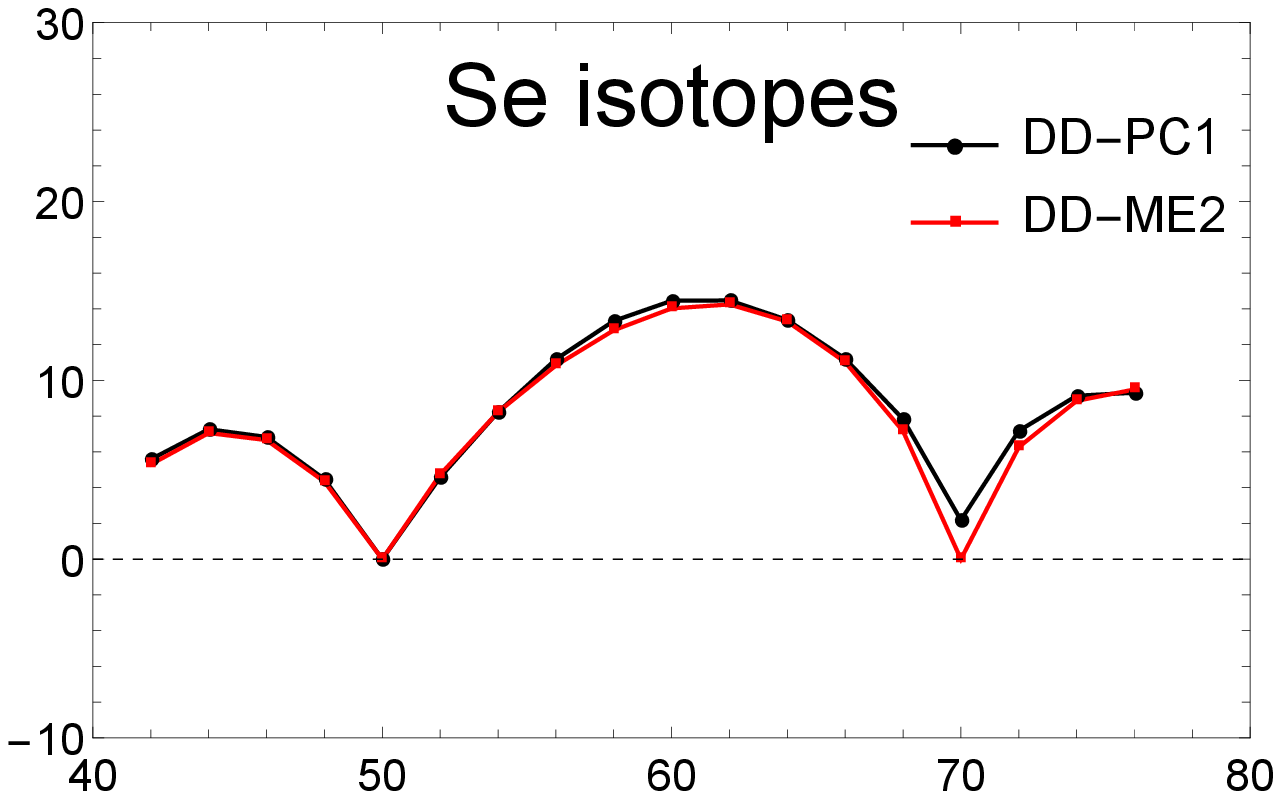}
 	\end{minipage}\hfill	
 	\begin{minipage}{0.5\textwidth}
 		\centering \includegraphics[scale=0.45]{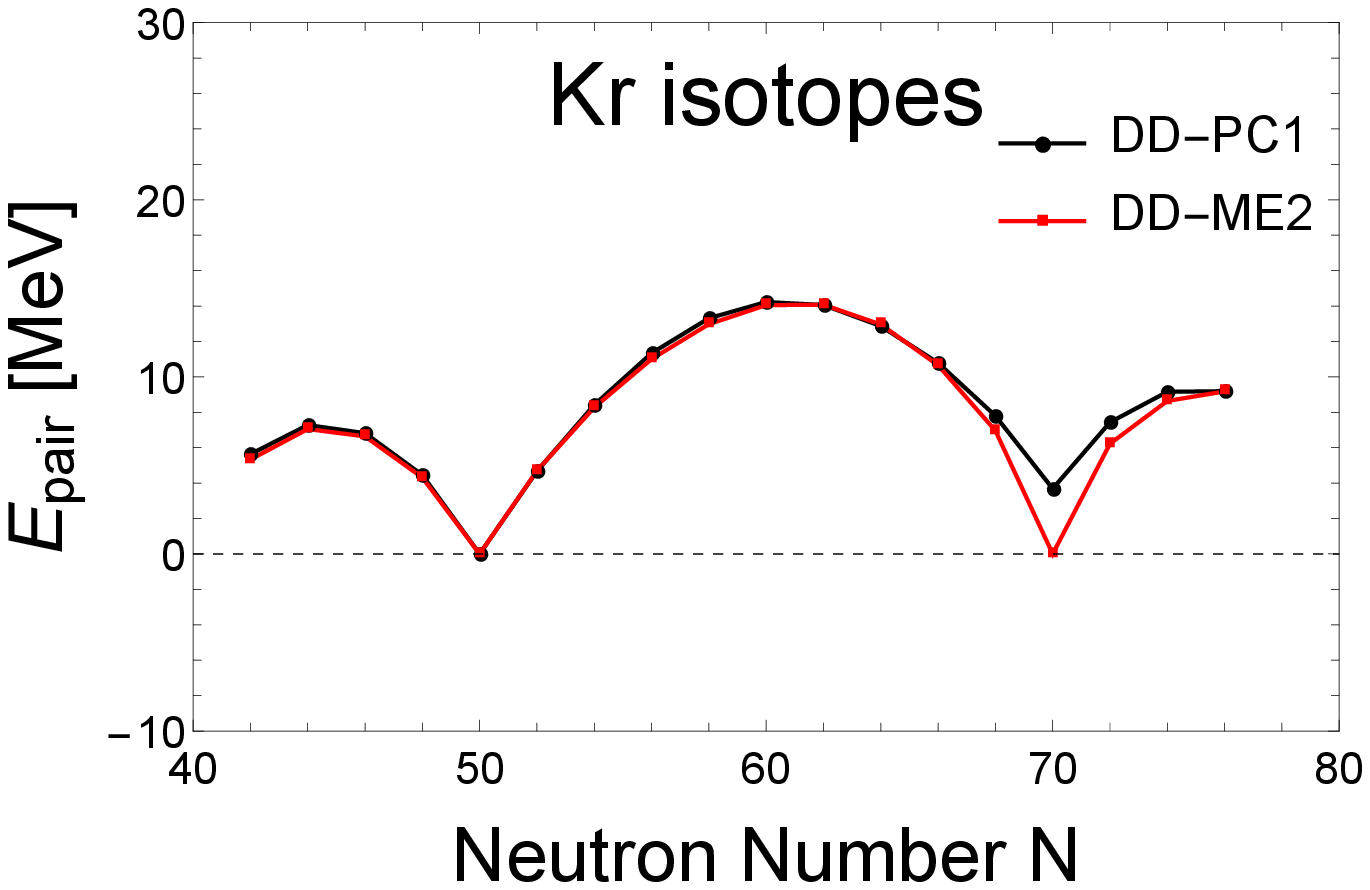}
 	\end{minipage}\hfill
 	\begin{minipage}{0.5\textwidth}
 		\centering \includegraphics[scale=0.45]{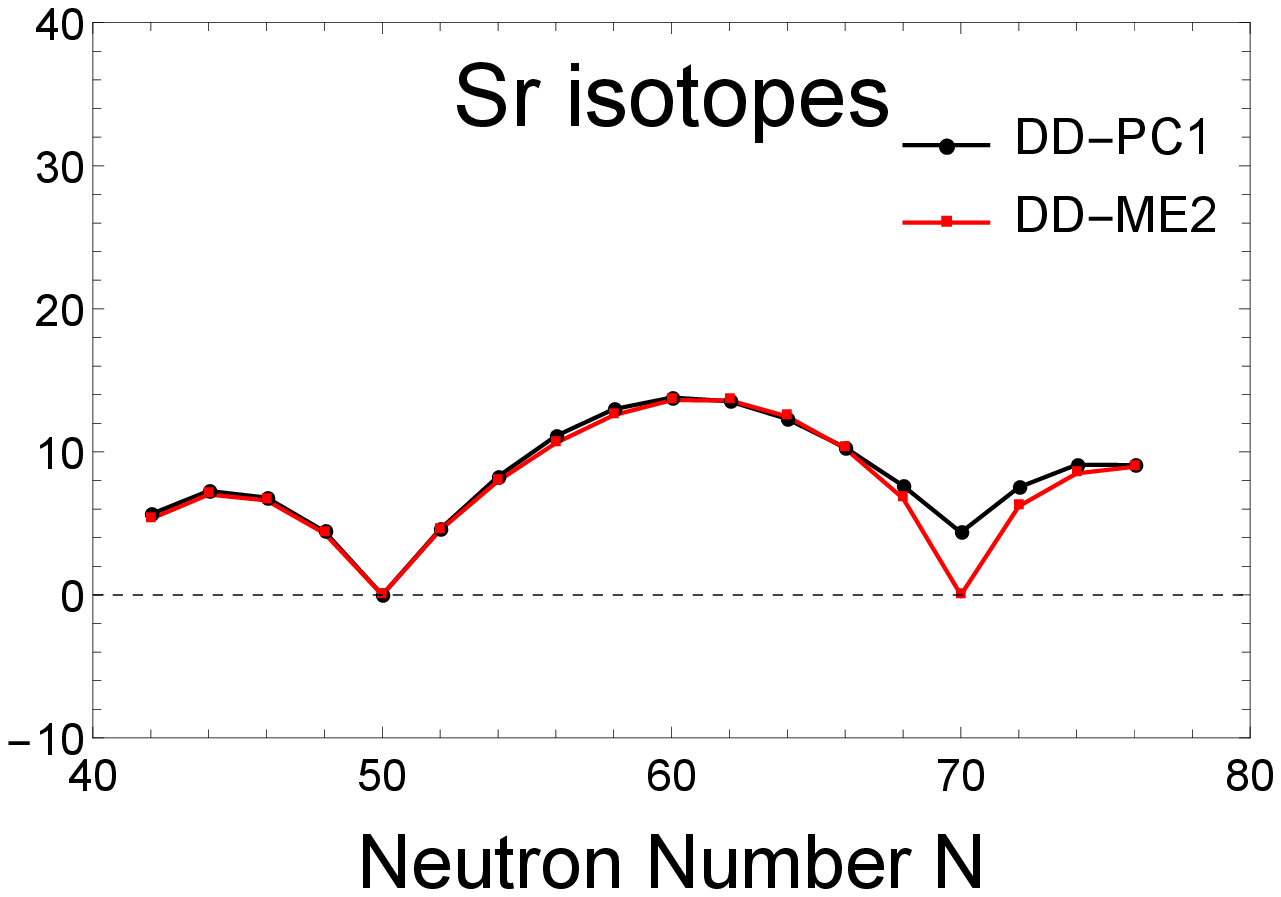}
 	\end{minipage}\hfill
 	\caption{The pairing energy for even-even Ge, Se, Kr and Sr isotopes calculated in RHB with DD-PC1 and DD-ME2 forces.}\label{EpairFig}
 \end{figure}

 The neutron pairing energy $E_{pair}$ vanishes when crossing a shell closure. This feature can be used to support our previous finding. In the RHB framework, this energy has been defined as
 \begin{equation}
 E_{pair}=\frac{1}{4}\sum\limits_{n_1 n'_1}\sum\limits_{n_2 n'_2}\kappa^*_{n_1n'_1}<n_1 n'_1 \mid V^{pp}\mid n_2 n'_2>\kappa_{n_2 n'_2}
 \end{equation}
 The matrix elements of the two-body pairing interaction and the original basis are noted $<n_1 n'_1 \mid V^{pp}\mid n_2 n'_2>$ and n, respectively. 
 In Fig. \ref{EpairFig}, we plot the variations of the neutron pairing energy $E_{pair}$ for the even-even $_{32}Ge$, $_{34}Se$, $_{36}Kr$ and $_{38}Sr$ isotopes calculated by RHB with the effective interactions DD-ME2 and DD-PC1. Both effective interactions are generally behaving in the same manner. In all cases, the neutron pairing energy vanishes at N=50 which establishes this classic magic number. At N=70, $E_{pair}$ also vanishes, except for Se, Kr and Sr where the minimum is to move slightly away from zero with decreasing proton number for DD-PC1 model. Despite that, both models reinforce our previous finding that N=70 is a new candidate neutron closure shell.\\
  
 \subsection{Potential energy surface}	
\begin{figure*}[tbh]
 	\centering  \includegraphics[scale=0.5]{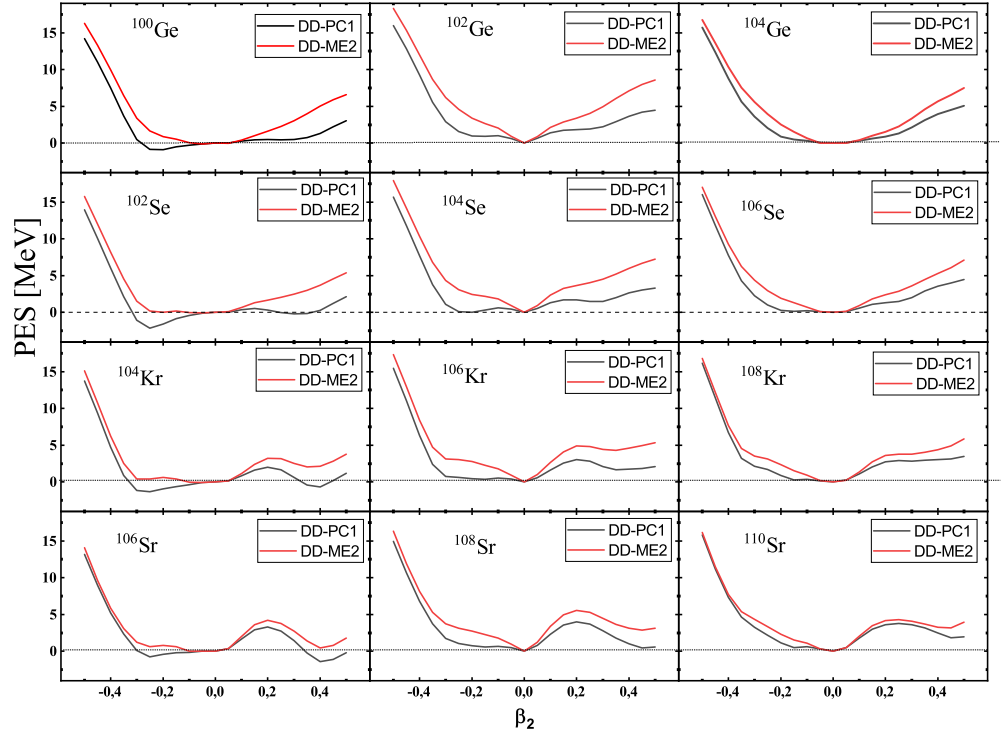}			
 	\caption{(Color online) The calculated potential surface energy for even-even Ge, Se, Kr and Sr nuclei calculated in RHB with DD-PC1 and DD-ME2 sets.}\label{ESurFig}
 \end{figure*}

 The potential energy surface (PES) represents the difference between the ground state energy and the energy at the deformation parameter $\beta_2$. Here, by using the triaxial DIRHB program, we have performed constrained calculations to extract the evolution of the  (PES) with the quadrupole moment $\beta_2$ which varies in a range from $\beta_2=-0.5$ to $\beta_2=0.5$ in steps of 0,05, as shown in Fig. \ref{ESurFig}. We have chosen PES=0 as a position to the ground state. The minimum of PES gives us an idea about the shape of nuclei; if it is attained at $\beta_2<0$, the nucleus has an oblate shape; if it is reached at $\beta_2>0$, the nucleus form is prolate; if it is reached at $\beta_2=0$, the nucleus is spherical.\\
 Aiming to corroborate the previous results that predicted N=70 as a neutron shell closure, we have plotted the Fig. \ref{ESurFig} which depicts the evolution of potential energy surface (PES) curves of Ge, Se, Kr and Sr isotones for N=70 (middle part of Fig. \ref{ESurFig}) and their neiboring isotones with N=68 (left part of Fig. \ref{ESurFig}) and N=72 (right part of Fig. \ref{ESurFig}) as functions of the quadrupole deformation $\beta_2$. From this figure, it is remarkable that both DD-ME2 and DD-PC1 sets  generally present the same behaviour. In all N=70 isotones, DD-ME2 and DD-PC1 show a sharp single minimum around the ground state, which confirms that the ground states of these isotones are spheric. Furthermore to their spheric ground states, the DD-PC1 potential curves for Se and Kr display  wide and  flat minima on the oblate side.
 For N=68 isotones, both models predict an oblate ground state for Se, Kr and Sr and an almost spherical, but slightly oblate ground state for Ge, Se, Kr and Sr N=72 isotones. For $^{100}Ge$, solely DD-ME2, predicts an oblate form, whereas DD-PC1 exhibits a flat minima toward the oblate region. 
  Again, with potential surface energy we arrive to prove the magicity in the investigated nuclei at neutron gap N=70.

 \subsection{Single particle energies}

\begin{figure}[tbh]
  		\centering \includegraphics[scale=0.45]{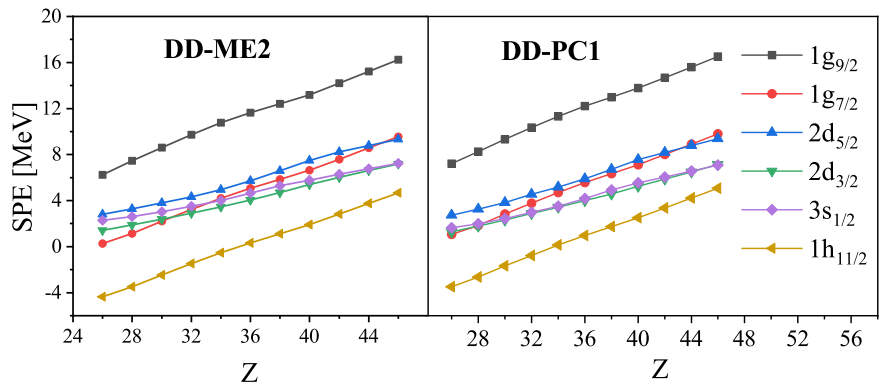} 
  	\caption{(Color online) The calculated single particle energies for the N=70 isotones $(26 \le Z \le 46)$ using the DD-PC1 and the DD-ME2 sets.}\label{SPEFig}
  \end{figure}

 To shed more light on the neutron shell closure at N=70, we have displayed in Fig. \ref{SPEFig} the neutron single particle energies (SPEs) as a function of the neutron number N around N=70 for $(26 \le Z \le 46)$, which is considered as a useful and sensitive observable for detecting shell gaps. 
  
 Let us start with SPEs of $3s_{1/2}$ and $1h_{11/2}$ states which correspond to the neutron shell gap N=70. In Fig. \ref{SPEFig}, we can see that the spacing between these states, with DD-ME2 set, is slightly greater than that obtained with DD-PC1 set. Despite this remark, the SPE in  both cases is significantly large which supports the shell closure at N=70 as predicted in the previous subsections. Furthermore, one can see that the N=70 gap is reduced when the proton number decreases, and became weaker up to $Z=38$. Moreover, $2d_{3/2}$ is quite close to $3s_{1/2}$, leaving the sub-shell at N=68 which induces the appearance of the bound gap  at N=70.
 
 We now turn to the study of the neutron shell gap N=50 that formed between $1g_{7/2}$ and $1g_{9/2}$ states. From Fig. \ref{SPEFig}, it is clearly seen that for both DD-ME2 and DD-PC1 sets, the large difference between $1g_{7/2}$ and $1g_{9/2}$ is basically unchanged when Z increases, which confirms the persistence of the classic shell closure N=50 in all investigated isotones. On the other hand, an inversion occurs between $1g_{7/2}$ and $2d_{5/2}$ states which initiates the emergence of a sub-shell at N=58.
 
 \section{Conclusion}\label{Sec5}
 To sum up, even-even neutron rich Ge, Se, Kr and Sr nuclei have been investigated in the relativistic Hartree-Bogoliubov (RHB) framework with DD-PC1 and DD-ME2 parametrizations. The Fermi levels, the two-neutron separation energies $S_{2n}$, the neutron shell gap $\delta_{2n}$, the neutron pairing gap $E_{pair}$, the potential energy surface (PES) and the energies of the canonical levels near the Fermi surface (SPE) reproduce, for all investigated nuclei, the well-known shell closure at N=50, in agreement with the results of Ref. \cite{Hakala}, and suggest the emergence of a new magic number at N=70, especially with DD-ME2. From analysing the evolution of the single-particle spectrum, it is found that the weaker gap at N=68 gave rise to the creation of new neutron shell closure just up, i.e. at N=70. The neutron drip-line for the four selected nuclei is speculated to be above N=70. Our (RHB) results are in good agreement with the available experimental data. Some discrepancies are shown between our calculations and FDRM predictions.



\begin{thebibliography}{99}
 			    \bibitem{Puldiner} B. S. Pudliner et al., Phys. Rev. lett. 74, 4396 (1995)
 			    \bibitem{Hagen}G. Hagen et al., Phys. Scr. 91, 063006 (2016)
 			    \bibitem{Barrett}B.R. Barrett et al., Prog. Part. Nucl. Phys. 69, 131 (2013) 
 			    \bibitem{Drut}J.E. Drut et al., Prog. Part. Nucl. Phys. 64, 120 (2010)
 			    \bibitem{Horoi} M. Horoi et al., Phys. Rev. C  50 (5), R2274 (1994)
 				\bibitem{Vautherin1969} D. Vautherin., and M. Veneroni, Phys. Lett. 29B, 203 (1969)
             	\bibitem{Vautherin1972} D. Vautherin, and D. M. Brink, Phys. Rev. C 5, 626 (1972)
             	\bibitem{Gogny} J. Dechargé, and D. Gogny, Phys. Rev. C 21 (4), 1568 (1980)
             	\bibitem{Walecka1974}  J. D. Walecka, Ann. Phys. (N.Y.) 83, 491 (1974)
             	\bibitem{Boguta1977} Boguta, J., and A. R. Bodmer, Nucl. Phys. A 292, 413 (1977)
                \bibitem{Niksic2011} T. Nik\v{s}i\'{c} , D. Vretenar, and P. Ring, Prog. Part. Nucl. Phys. 66 519 (2011)
             	\bibitem{Agbemava2015} S. E. Agbemava, A. V. Afanasjev, T. Nakatsukasa, and P. Ring, phys. Rev. C 92, 054310 (2015)
             	\bibitem{Bassem} Y. El Bassem, M. Oulne,  Nuclear Physics A, 987, 16-28 (2019)
             	\bibitem{Lie} J. J. Li et al., Phys. Lett. B 753 : 97-102 (2016).
             	\bibitem{ELADRI} M. El Adri, M. Oulne, Eur. Phys. J. Plus 135, 268 (2020)
             	\bibitem{Bhattacharya} M. Bhattacharya, G.Gangopadhyay, Phys. Rev. C, 72(4), 044318 (2005)
             	\bibitem{Hakala} J. Hakala et al., Phys. Rev. Lett. 101, 052502 (2008)
               	\bibitem{Nakamura} T. Nakamura et al., Prog. Particle Nuclear Phys. 97, 53 (2017)
               	\bibitem{Motobayashi} T. Motobayashi, et al., Phys. Lett. B 346 (1995) 9
               	\bibitem{Kanungo} R. Kanungo, Phys Scr. T. 152, 014002 (2013)
               	\bibitem{Grawe} H. Grawe et al., Rep. Prog. Phys. 70, 1525 (2007)
               	\bibitem{Niksic2008} T. Nik\v{s}i\'{c}, D. Vretenar, and P. Ring, Phys. Rev. C 78 034318 (2008)
             	\bibitem{Lalazissis2005} G. A. Lalazissis, T. Nik\v{s}i\'{c} , D. Vretenar, and P. Ring, Phys. Rev. C 71 024312 (2005)
             	\bibitem{Niksic2014} T. Nik\v{s}i\'{c}, N. Paar, D. Vretenar et P. Ring, Comput. Phys. Commun 185, 1808-1821 (2014)
                \bibitem{Walker1986} B. B. Serot and J. D. Walecka, Adv. Nucl. Phys. 16 1 (1986)
                \bibitem{Typel1999} S. Typel and H. H. Wolter, Nucl. Phys. A 656 331 (1999)
 				\bibitem{Tian2009} Y. Tian, and Z. Y. Ma, P. Ring, Phys. Lett. B 676 44 (2009)
 				\bibitem{Wang} M. Wang, G. Audi, A. H. Wapstra and al. The Ame2016 atomic mass evaluation. Chinese Physics C, 41, 030003 (2017)
 		    	\bibitem{Moller} P. M\"oller, A. J. Sierka, T. Ichikawab and H. Sagawa,	Atomic Data and Nuclear Data Tables 109, 1-204 (2016)
 		    	\bibitem{Bender}M. Bender, et al. Phys, Rev, C 80(6), 064302 (2009)
 		    	\bibitem{Vretenar} D. Vretenar et. al., Physics reports, 409(3-4), 101-259 (2005)	
 			\end{thebibliography}
 
 \end{document}